\documentclass[%
 aip,
 amsmath,amssymb,
 reprint,%
]{revtex4-1}

\usepackage{graphicx}
\usepackage{dcolumn}
\usepackage{bm}
\usepackage{xcolor}
\usepackage[utf8]{inputenc}
\usepackage[T1]{fontenc}
\usepackage{mathptmx}
\usepackage{amssymb}
\usepackage{amsmath}

\newcommand{\pfr}[2]{\ensuremath{\frac{\partial #1}{\partial #2}}}
 
\newcommand{\dd}{\mathrm{d}}
\newcommand{\Sch}{\textit{Sc}}

\newcommand{\red}[1]{\textcolor{black}{#1}}

\newcommand{\beq}{\begin{equation}}
\newcommand{\eeq}{\end{equation}}

\begin{document}

\preprint{AIP/123-QED}

\title[Solute disperion in straining flows and boundary layers]{Dispersion of solute in straining flows and boundary layers}

\author{Prabakaran Rajamanickam}
 
 \email{pzr0023@auburn.edu}
 \affiliation{ 
Department of Aerospace Engineering, Auburn University, Auburn, AL 36849, USA 
}%
 

\date{\today}

\begin{abstract}
Solute dispersion due to an instantaneously released source in steady, laminar, axisymmetric flows with an axial inflow and radial outflow is investigated analytically. Attention is given to large-time characteristics of dispersion, where the concentration reduces in proportion to $e^{-2\lambda \tau}$, where $\lambda$ is an eigenvalue that depends on the axial inflow and $\tau$ is the time measured in units of axial-diffusion times. \red{Prospects of some other flows are also considered.}
\end{abstract}

\maketitle
 
The simplest example of shear-induced dispersion arising in unidirectional flows, a case addressed by G. I. Taylor in the early 1950s,\cite{taylor1953dispersion,taylor1954dispersion} illustrates how at times large compared to the radial-diffusion times, the combined action of axial convection and radial diffusion results in an axial-diffusion process, with an enhanced diffusivity. Although this is the first important step taken by Taylor to quantify dispersion processes, when the flow is not \red{nearly unidirectional}, the resulting dispersion processes become quite complicated due to the strong interactions between convection and diffusion. A generalized theory for arbitrary flow fields is a formidable task. In this paper, we aim to examine the dispersion process in some axisymmetric flow fields described below, that appears to merit consideration from the viewpoint of engineering applications.

\red{We use cylindrical coordinates $(r,\theta,z)$ with corresponding velocity components $(v_r,v_\theta,v_z)$.} The radial and axial velocity components of the form
\begin{equation}
    v_r = krF'(\eta), \quad v_z = -2 \sqrt{k\nu}F(\eta) \label{vrvz}
\end{equation}
where $\eta=z/\sqrt{\nu/k}$, $\nu$ is the kinematic viscosity of the fluid and $k$ is a constant parameter characterising the flow field, accommodate a wide variety of solutions of the incompressible Navier-Stokes equations. This velocity form corresponds to supposing that with both $F>0$ and $F'>0$ in the upper-half plane $z>0$, the inflow motion is along $z$-axis directed to the \red{stagnation} plane $z=0$, while the fluid leaves the region of interest radially outwards to infinity. 

An elementary form for the function 
\begin{equation}
    F(\eta)=\eta, \label{counterflow}
\end{equation}
corresponds to axisymmetric counterflows, where two opposing round jets are allowed to impinge on the stagnation-plane. This flow set-up is ubiquitous in combustion experiments,\cite{niemann2015accuracies} in studying the effects of strain-rates $k$ on flames. When a round jet is allowed to impinge on a solid wall, the flow field obtained is usually referred to as the Homann flow, for which $F(\eta)$ satisfies the equation
\begin{equation}
    F''' + 2FF'' - F'^2 + 1=0 \label{homann}
\end{equation}
subjected to conditions $F(0)=F'(0)=F'(\infty)-1=0$. Another interesting solution is due to a flow induced by rotating disks, widely known as the von k\'arm\'an swirling flow, in which case, the parameter $k$ denotes the rotation rate of the disk and the function $F$ satisfies the coupled equation
\begin{equation}
    F''' + 2FF'' - F'^2 + G^2=0, \quad G'' + 2 FG' - 2F'G=0, \label{karman}
\end{equation}
conditioned by $F(0)=F'(0)=F'(\infty)=G(0)-1=G(\infty)=0$. The function $G(\eta)$ determines the azimuthal motion $v_\theta=kr G(\eta)$. Additional solutions consistent with equations~\eqref{vrvz} are obtained by modifying the boundary conditions viz., conditions pertaining to stretching or shrinking walls, rotational inflows\cite{agrawal1957new,davey1963rotational} and so on, as has been reviewed in these references.\cite{drazin2006navier,wang2011review} \red{Recent interests in flows of the form~\eqref{vrvz} with applications directed towards micropolar fluids~\cite{nazar2004stagnation,ishak2010stagnation}, nanofluids~\cite{mustafa2011stagnation,farooq2016mhd}, MHD flows~\cite{mahapatra2001magnetohydrodynamic}, flows in porous media~\cite{wu2005stagnation}, etc., further enlarge the range of usefulness of the current investigation.}

In those flows described above, a concentrated soluble matter of concentration $c$ will be released at the origin (the analysis to be presented can be extended easily to \red{one where the concentration is released at an arbitrary location}) at a prescribed time, say at $t=0$. Further evolution of this concentrated cloud is to be investigated for times large compared to the diffusion times. Since the initial condition is axially symmetric (in fact, spherically symmetric) and all the flow fields considered are inherently axisymmetric, only axisymmetric solutions of $c$ will be sought. The azimuthal velocity, if present, will then have no effect on $c(r,z,t)$. The equation governing the concentration $c$ can be written in the form
\begin{equation}
   \pfr{c}{t} + v_r \pfr{c}{r}  + v_z \pfr{c}{z} = D\left[\frac{1}{r}\pfr{}{r}\left(r\pfr{c}{r}\right) +\pfr{^2c}{z^2}\right]  \label{maindim}
\end{equation}
where $D$ denotes the molecular diffusivity of the solute, assumed here to be constant. The solution of the above equation satisfy the boundary conditions 
\begin{align}
     \pfr{c}{r}=0 &\quad \text{at}\quad r=0,\\
     \pfr{c}{z}=0 &\quad \text{at}\quad z=0,\\
     c=0 &\quad \text{as} \quad r^2+z^2\rightarrow \infty
\end{align}
and the initial condition 
\begin{equation}
    t=0: \qquad  c = \frac{M}{\pi r}\delta(r) \delta(z) \label{maininit}
\end{equation}
where $M$ has the units of that of the concentration multiplied by the volume that the solute occupies and $\delta$ is the Dirac delta function. At this point, it is worth mentioning that the associated problems in two-dimensional planar geometries were addressed by Chatwin.\cite{chatwin1974dispersion} However, the large-time solutions constructed there require revisions, as we shall explore at the end.

To facilitate the analysis, the equations described in~\eqref{maindim}-\eqref{maininit} will be non-dimensionalized suitably by introducing the rescaled coordinates $\tau=kt$, $\xi=r/\sqrt{\nu/k}$ and $\eta=z/\sqrt{\nu/k}$ and a rescaled concentration $C= (c/M) (\nu/k)^{3/2}$, to yield
\begin{equation}
   C_\tau + \xi F' C_\xi - 2 F C_\eta =  \Sch^{-1} [(\xi C_\xi)_\xi/\xi + C_{\eta\eta}]  \label{maineq}
\end{equation}
where $\Sch= \nu/D$ is the Schmidt number of the solute. Accordingly, the boundary conditions become
\begin{align}
     C_\xi=0 &\quad \text{at}\quad \xi=0,  \label{rbc}\\ 
     C_\eta=0 &\quad \text{at}\quad \eta=0,  \label{zbc} \\
     C=0 &\quad \text{as} \quad \xi^2+\eta^2\rightarrow \infty, \label{infbc}
\end{align}
whereas the initial condition takes the form
\begin{equation}
    t=0: \qquad  C = \frac{1}{\pi\xi} \delta(\xi)\delta(\eta). \label{initial}
\end{equation}

First, we shall consider the axisymmetric counterflows with $F$ given by~\eqref{counterflow}, for which an explicit solution can be found. \red{In this particular case where $F=\eta$}, $\Sch$ can be set to unity, since it can be removed from~\eqref{maineq}-\eqref{initial} by the transformation $(\xi,\eta,C)\rightarrow \sqrt\Sch(\xi,\eta,\Sch\, C)$; in other words, by scaling the dimensional variables using $D$ \red{as the characteristic diffusivity} in place of $\nu$, since the velocity components here are independent of $\nu$. The explicit solution is simply the product of two fundamental solutions each containing only one independent coordinate and the time variable $\tau$,
\begin{equation}
    C(\xi,\eta,\tau) = \frac{\exp\{-\xi^2/[2(e^{2\tau}-1)]-\eta^2/(1-e^{-4\tau})\}}{\pi (1-e^{-2\tau})\sqrt{\pi(e^{4\tau}-1)}} .  \label{explicit}
\end{equation}
For small times, $\tau\ll 1$, the asymptotic expansion is
\begin{equation}
    C\rightarrow \frac{2e^{-\frac{\xi^2+\eta^2}{4\tau}}}{(4\pi\tau)^{3/2}} \left( 1 + \frac{\xi^2-2\eta^2}{4} + \cdots \right). \label{smallt}
\end{equation}
This implies that the leading behavior is that of the spherically-symmetric diffusive spreading of an instantaneous source. This result should be obvious since for small times, the diffusion process spreads an initially concentrated cloud with a characteristic thickness $\sqrt\tau$, meanwhile both the radial and axial convection are able to convect the solute only to a distance of order $\tau$ from the origin. Therefore the diffusive behaviour for small times becomes independent of $F(\eta)$. The first-order correction in~\eqref{smallt} arising due to convection already destroys the spherical symmetry.

On the other hand, the large-time ($\tau \gg 1$) behavior,
\begin{equation}
    C\rightarrow  \frac{1}{\pi \sqrt\pi }\exp\left[-2\tau-\frac{\xi^2}{2e^{2\tau}}-\eta^2\right] + \cdots \label{larget}
\end{equation}
needs an explanation. The above expansion implies that the concentration decays like $C\sim e^{-2\tau}$ with a typical radial width $\xi\sim e^\tau$. At times $\tau \gg 1$, the competing balance between convection and diffusion in the axial direction would keep the concentration in an axial extent of order $\eta\sim 1$. Thus the non-dimensional scale $\eta$ is the requisite one. However, as the cloud spreads away from the origin, the radial convection, proportional to $\xi$ becomes dominant compared to the radial diffusion. As $\tau$ becomes large, the solute will then have spread to a radial extent that is determined from the balance
\begin{equation}
    C_\tau + \xi C_\xi = 0. \label{convection}
\end{equation}
The solution to the foregoing equation is $C=C(\xi e^{-\tau})$. The function $C$ can be evaluated from solutions found at $\tau\sim 1$, where because of the combined action of convection and diffusion in both directions, the solute will be dispersed such that $C$ is of order unity for $\xi<1$ and of negligible order for $\xi>1$. This information about the shape of $C$ will be transmitted to large times without much variations to it, although the magnitude of $C$ is not expected to stay the same. However, \red{we can assert that} the ratio $C(\xi e^{-\tau})/C(0)$ will still be an order-unity quantity for $\xi e^{-\tau}<1$ and negligible for $\xi e^{-\tau}>1$. It is clear then that the \red{radial} scale we are looking for is $\xi\sim e^\tau$. As the solute convects to these large radial distances, the overall magnitude of $C$ will drop quickly due to the conservation of matter. The characteristic peak value of $C$ is then inversely proportional to a characteristic volume occupied by the solute at times $\tau\gg 1$, i.e.
\begin{equation}
    C \sim\frac{1}{\xi^2\eta}\sim e^{-2\tau}.
\end{equation}
 As we can see, the scaling arguments are consistent with the asymptotic behavior~\eqref{larget}, \red{obtained from the explicit solution~\eqref{explicit}. Note that although the balance~\eqref{convection} does provide a correct scale for the radial coordinate, it is not uniformly valid; the radial-diffusion term is no longer negligible when $\xi\sim 1$.} 
 
 \red{Extending these scaling laws over to the case where $F\neq \eta$, one may find the characteristic length scales as $\eta\sim 1$, $\xi\sim e^{-F'\tau}$ and the characteristic value for the concentration as $C\sim e^{-2F'\tau}$. However, this would imply different decay rates for $C$ along $\eta$, In particular, a zero decay rate at the wall and thereby violating the conservation of matter since the initially concentrated cloud has now spread to a radial extent $e^{-\mathrm{max}(F')\tau}$. Now that the equations~\eqref{maineq}-\eqref{initial} does provide no explicit solution, the correct decay rate for $C$ will be derived by analyzing the large-time asymptotic characteristics of the governing equation. Before doing that, first let us} investigate Chatwin's method.\cite{chatwin1974dispersion} An equation similar to~\eqref{maineq} appears for the corresponding two-dimensional problem. Guided by the solution obtained for the counterflows, Chatwin introduced the ansatz
 \begin{equation}
     C = A(\xi,\tau) e^{-2\Sch \int F(u)\, \dd u} \label{chatwin}
 \end{equation}
 with $A$ satisfying the equation $A_\tau + F' (\xi A)_\xi +A F' = \Sch^{-1} (\xi A_\xi)_\xi/\xi$. Since $A$ is a function only of $\xi$ and $\tau$ and the preceding equation contains a function of $\eta$, no meaningful solutions can be obtained unless $F'$ is a constant. Thus, Chatwin proposed to replace $F'$ with its average value (suitably defined) so that an expression similar to the asymptotic expansion~\eqref{larget} can be derived for the general case where $F\neq \eta$. However, numerical integrations of~\eqref{maineq}-\eqref{initial} do not seem to agree with these predictions.

The failure of the ansatz~\eqref{chatwin} is due to the assumption that the time dependence of $C$ is coupled with the independent variable $\xi$, while on the other hand, we can already see from~\eqref{larget} that the $e^{-2\tau}$ decay for $C$ has been originated from the fundamental solution corresponding to the independent variable $\eta$. In order to find the correct ansatz, first let us assume that
\begin{equation}
    C_\tau + \xi F' C_\xi = \Sch^{-1}(\xi C_\xi)_\xi/\xi.
\end{equation}
With errors of order $e^{-F'\tau}$, the solution to  the foregoing equation is
\begin{equation}
    C\sim \exp\left(- \frac{\Sch F' \xi^2}{2e^{2F'\tau}}\right). \label{Cxi}
\end{equation}
This suggests us to consider the ansatz
\begin{equation}
    C=A(\eta,\tau) \exp\left(- \frac{\Sch F' \xi^2}{2e^{2F'\tau}}\right),
\end{equation}
so that the time decay for $C$ is determined by the fundamental solution pertaining to $\eta$. We now take the above assumed form into~\eqref{maineq} and collect terms of different orders corresponding to the limit $\tau\rightarrow \infty$. At leading order, we find
\begin{equation}
    A_\tau - 2 F A_\eta = \Sch^{-1} A_{\eta\eta} \label{Aeq}
\end{equation}
The error introduced in writing this equation of order $\tau^2 e^{-F'\tau}$ is larger than the error magnitude introduced in obtaining~\eqref{Cxi}. 

Equation~\eqref{Aeq} admits solutions of the form $A(\eta,\tau)=e^{-2\lambda \tau} g(\eta)$, where
\begin{equation}
    \Sch^{-1} g'' + 2 F g' + 2 \lambda g =0 \label{eigen}
\end{equation}
with boundary conditions $g'(0)=g(\infty)=0$. This is a Sturm-Liouville problem as can be seen by rewriting the equation as
\begin{equation}
    \left(e^{2 S\int F\,\dd u}g'\right)' + 2 S e^{2 S\int F\,\dd u} \lambda g =0.
\end{equation}
It can be shown that the eigenvalues are always positive such that $C$ decays with time.
\begin{figure}
    \centering
    \includegraphics[width = 0.9\linewidth]{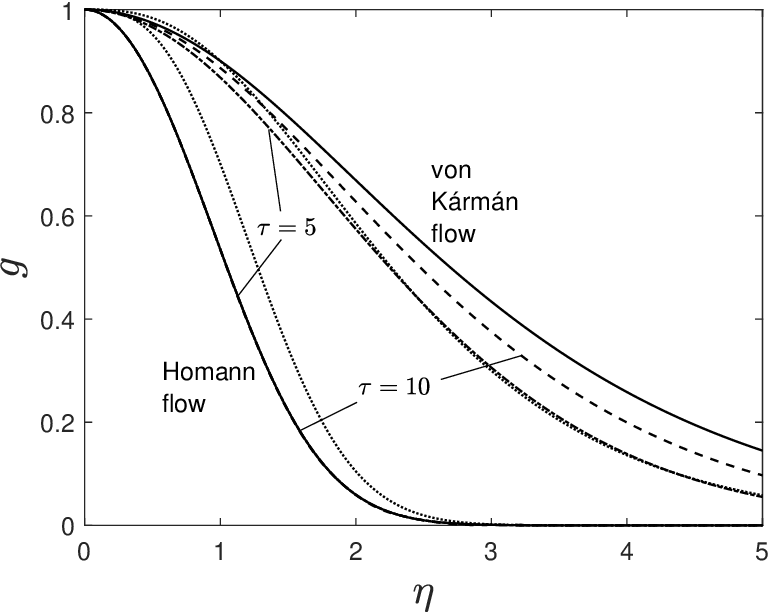}
    \caption{The eigenfunctions $g(\eta)$ are represented by the solid lines. The dash-dashed and dashed curves correspond to $C(0,\eta,\tau)/C(0,0,\tau)$, computed from~\eqref{maineq} for $\tau=5$ and $\tau=10$, respectively. The dotted curves correspond to Chatwins approximation~\eqref{chatwin}. All curves are calculated for $\Sch=1$.}
    \label{fig:eig}
\end{figure}
Since we are looking for solutions as $\tau\rightarrow\infty$, only the smallest eigenvalue $\lambda$ and \red{the corresponding} eigenfunction $g(\eta)$ are required. Then the large-time solution for $C$ is given by
\begin{equation}
    C = K\, g(\eta) \exp\left[-2\lambda \tau - \frac{\Sch F' \xi^2}{2e^{2F'\tau}} \right] + \cdots \label{largetime}
\end{equation}
where the constant $K$ can be determined from condition
\begin{equation}
    \int_0^\infty \int_0^\infty \xi C\, \dd\xi\, \dd\eta = \frac{1}{2\pi}.
\end{equation}
\red{Note that the accurate determination of $K$ requires more terms in the expansion~\eqref{largetime}.}

Since for axisymmetric counterflows, \red{the Schmidt number can be set equal to one,} the eigenvalue equation becomes $g'' + 2 \eta g' + 2 \lambda g =0$. Introducing the substitution $g(\eta)=e^{-\eta^2}h(\eta)$ reduces it to the Hermite equation $h''-2\eta h' + 2(\lambda-1) h=0$. As we know, finite solutions to the Hermite equation exist only for positive integral (and zero) eigenvalues.\cite{landau2013quantum} The eigenfunctions are, of course, Hermite polynomials; the first of which corresponding to the eigenvalue $\lambda-1=0$ is unity provided $h$ \red{has been} normalized appropriately. \red{As we required, this result for axisymmetric counterflows is in confirmation with equation~\eqref{larget}.} For the general case, the eigenvalue equation~\eqref{eigen} has to be solved numerically. With $\Sch=1$, the smallest eigenvalue for the Homann flow is $\lambda =0.5663$ and for the von k\'arm\'an flow, it is $\lambda=0.1046$.

In figure~\ref{fig:eig}, we compare the eigenfunction $g(\eta)$ calculated from~\eqref{eigen} with the concentration profiles along the axis calculated from~\eqref{maineq}-\eqref{initial} at two different values of $\tau$. All figures are scaled with respect to their values found at $\eta=0$, thereby making the comparison easier without having to determine $K$. For the two chosen values of $\tau$, the eigenfunctions predict the axial variations of $C$ with greater accuracies for the Homann flows \red{(in the figure, the actual profiles are indistinguishable from the eigenfunction)} than the von K\'arm\'an swirling flows because the error magnitude $\tau^2 e^{-F'\tau}$ in von K\'arm\'an flows is comparatively larger. \red{Plotted in the same figure are the functions $\exp(-2\int_0^\eta F(u)\,\dd u)$, corresponding to Chatwin's approximations. As can be seen from the figure, the overlapping of Chatwin's approximation with the full numerical solution for the von k\'arman flows with $\tau=5$ in the range $\eta>2$ is accidental as can be inferred from the computations for $\tau=10$.}

Thus far we have considered fluid motions with axial inflow $F>0$ and radial outflow $F'>0$. The opposite motion where we have a radial inflow $F'<0$ and an axial outflow $F<0$ finds applications in stretched vortex lines. For instance, Burgers vortex is a simplest example where
\begin{equation}
    F(\eta) = -\eta
\end{equation}
describes the motion in the axial plane, for which an explicit solution can be found. The solution is given by
\begin{equation}
    C(\xi,\eta,\tau) = \frac{\exp\{-\xi^2/[2(1-e^{-2\tau})]-\eta^2/(e^{4\tau}-1)\}}{\pi (e^{2\tau}-1)\sqrt{\pi(1-e^{-4\tau})}}, \label{explicit2}
\end{equation}
whose asymptotic behavior for large $\tau$ becomes
\begin{equation}
    C\rightarrow \frac{1}{\pi \sqrt\pi }\exp\left[-2\tau-\frac{\xi^2}{2}-\frac{\eta^2}{e^{4\tau}}\right] + \cdots \label{larget2}
\end{equation}
As expected, the concentration is confined to a radial extent $\xi$ of order unity and extended in the axial direction of order $\eta\sim e^{2\tau}$. \red{Explicit solutions can be constructed for the general three-dimensional linear shear flows from the product of three fundamental solutions each corresponding to three different Cartesian coordinates.}

In concluding this paper, we extend the large-time asymptotic solution to two-dimensional planar flows with Cartesian velocity components given by
\begin{equation}
    v_x = k xF'(\eta), \quad v_z = -\sqrt{k\nu} F(\eta).
\end{equation}
Assuming both $F>0$ and $F'>0$, the large-time solution is given by
\begin{equation}
    C = K\, g(\eta) \exp\left[-\lambda \tau - \frac{\Sch F' \xi^2}{2e^{2F'\tau}} \right] + \cdots
\end{equation}
where $\tau=kt$, $\xi=x/\sqrt{\nu/k}$ and $\eta=z/\sqrt{\nu/k}$. The function $g(\eta)$ satisfies the equation
\begin{equation}
    \Sch^{-1} g'' + F g' + \lambda g = 0.
\end{equation}
subjected to the condition $g'(0)=g(\infty)=0$.

The author would like to acknowledge Adam D. Weiss for his interesting comments.

The data that supports the findings of this study are available within the article.

\bibliography{aipsamp}

\end{document}